\documentclass{iopart}
\begin{document}
\input epsf

\title[Monitoring of the thermal neutron flux in the LSM underground laboratory]{Monitoring
of the thermal neutron flux in the LSM underground laboratory}

\newcommand{\jinr}{\affiliation{Laboratory of Nuclear Problems,
Joint Institute for Nuclear Research, Dubna, 141980, Russia}}

\author{
S.~Rozov$^1$,
E.~Armengaud$^2$,
C.~Augier$^3$,
L.~Berg\'e$^4$,
A.~Benoit$^5$,
O.~Besida$^2$,
J.~Bl\"{u}mer$^{6,7}$,
A.~Broniatowski$^4$,
V.~Brudanin$^1$,
A.~Chantelauze$^7$,
M.~Chapellier$^4$,
G.~Chardin$^4$,
F.~Charlieux$^3$,
S.~Collin$^4$,
O.~Crauste$^4$,
M.~De~Jesus$^3$,
X.~Defay$^4$,
P.~Di~Stefano$^3$,
Y.~Dolgorouki$^4$,
J.~Domange$^4$,
L.~Dumoulin$^4$,
K.~Eitel$^7$,
D.~Filosofov$^1$,
J.~Gascon$^3$,
G.~Gerbier$^2$,
M.~Gros$^2$,
M.~Hannawald$^2$,
A.~Juillard$^{3,4}$,
H.~Kluck$^7$,
V.~Kozlov$^7$,
R.~Lemrani$^2$,
A.~Lubashevskiy$^1$,
C.~Marrach$^4$,
S.~Marnieros$^4$,
X-F.~Navick$^2$,
C.~Nones$^4$,
E.~Olivieri$^4$,
P.~Pari$^8$,
B.~Paul$^2$,
V.~Sanglard$^3$,
S.~Scorza$^3$,
S.~Semikh$^1$,
M-A.~Verdier$^3$,
L.~Vagneron$^3$
and E.~Yakushev$^1$\footnote{Corresponding author.}
}

\address{$^1$ Laboratory of Nuclear Problems, JINR, Joliot-Curie 6, 141980 Dubna, Moscow region, Russia}
\address{$^2$ CEA, Centre d'Etudes Saclay, IRFU, 91191 Gif-Sur-Yvette Cedex, France}
\address{$^3$ Universite de Lyon, F-69622, Lyon, France; Universite de Lyon 1, Villeurbanne; CNRS/IN2P3, Institut de Physique Nucleaire de Lyon, France}
\address{$^4$ Univ. Paris 11 CNRS, IN2P3, Centre Spec Nucl and Spect Masse, 91405 Orsay, France}
\address{$^5$ CNRS-Neel, 25 Avenue des Martyrs, 38042 Grenoble cedex 9, France}
\address{$^6$ Karlsruhe Institute of Technology, Institut f\"{u}r Experimentelle Kernphysik, Gaedestr. 1, 76128 Karlsruhe,
Germany}
\address{$^7$ Karlsruhe Institute of Technology, Institut f\"{u}r Kernphysik, Postfach 3640, 76021 Karlsruhe,
Germany}
\address{$^8$ CEA, Centre d'Etudes Saclay, IRAMIS, 91191 Gif-Sur-Yvette Cedex, France}

\ead{yakushev@jinr.ru}

\begin{abstract}
This paper describes precise measurements of the
thermal neutron flux in the LSM underground laboratory in proximity
of the EDELWEISS-II dark matter search experiment
together with short measurements at various other locations.
Monitoring of the
flux of thermal neutrons is accomplished using a mobile detection
system with low
background proportional counter filled with $^3$He. On average 75
neutrons per day are detected with a background level below 1 count
per day (cpd). This provides a unique possibility of a day by day
study of variations of the neutron field in a deep underground site.
The measured average 4$\pi$ neutron flux per cm$^{2}$ in the proximity of EDELWEISS-II
is $\Phi_{MB}=3.57\pm0.05^{stat}\pm0.27^{syst}\times 10^{-6}$~neutrons/sec.
We report the first experimental observation that the point-to-point
thermal neutron flux at LSM varies by more than a factor two.
\end{abstract}


\submitto{\JPG}


\section{Introduction}

The Laboratoire Souterrain de Modane (LSM)~\cite{lsm} is an
underground laboratory located in the Fr\'{e}jus tunnel connecting
France and Italy. It has an average rock overburden corresponding to $\sim$4850~mwe
(meter water equivalent)~\cite{ref_gabriel}. This reduces the muon flux by more than 6
orders of magnitude and the neutron flux by 4 orders of magnitude
compared to sea level. A principal source of remaining neutrons is
natural radioactivity in rock and all other materials located in
the laboratory, via spontaneous fission (SF) and ($\alpha$,n)
reactions. Many experiments are located at LSM, the two largest being
the EDELWEISS-II~\cite{edelweiss} direct Dark Matter search and the
NEMO-3~\cite{nemo3} experiment searching for neutrinoless double
beta decay ($0\nu\beta\beta$). An unbiased interpretation of results
from these experiments requires a detailed understanding of
all background sources. In the near future, it is planned to enlarge
LSM to host the next generation of experiments searching for Dark
Matter and $0\nu\beta\beta$. These will require further
background reductions and a precise knowledge of the remainder.
Thermal neutrons are not a background of concern for present-day
experiments, but may become so for future ones
where neutron activation of construction materials or
the detectors themselves may be an issue. For a correct
interpretation of results of dark matter experiments not only the
background rate but also its temporal variations are of critical
importance. The EDELWEISS-II collaboration performs a wide
variety of measurements of neutrons in the proximity of the
experimental setup together with Monte Carlo studies (MC)~\cite{lemrani0}.
Experimental studies include: a) the continuous monitoring of fast
neutrons since the start of data taking with EDELWEISS-II in 2006;
b) the measurement of fast neutrons produced by cosmic
muons in the vicinity of the EDELWEISS-II muon veto system;
c) the present measurement of thermal neutrons; and d) a feasibility
test of cryogenic detectors made of light
targets to study the fast neutron background~\cite{pcf}.

In this letter we present the study devoted to the measurement and
continuous monitoring of the thermal neutron flux at LSM, especially
in the proximity of the EDELWEISS-II setup.

\section{Thermal neutron detector \label{section_det}}

The detection of thermal neutrons is provided by a proportional
counter tube filled with $^3$He gas via the capture reaction
n$+^3$He$\to$T$+$p ($Q=0.764$~MeV). The cross section for thermal
neutrons on $^3$He is $\sigma=5333\pm 7$~barn~\cite{table_of_isotop}. The
CHM-57 counter~\cite{chm57} used in the setup has a working length
of 860~mm with an internal diameter of 31~mm. The counter is filled
with 400~kPa of $^{3}$He and 500~kPa of $^{40}$Ar as working gas.
The region of interest for thermal neutron detection are
ionization signals with energies around $Q$.
In proportional counters, the main background in this energy range
arises from
$\alpha$-decays of U and Th progenies in the walls of the detector.
To reduce this background, the CHM-57 counter is covered inside by a
50-60~$\mu$m thick layer of Teflon followed by a 1~$\mu$m layer of
electrolytic copper.

\begin{figure}
\begin{center}
\noindent\epsfxsize=0.75\textwidth \epsfbox{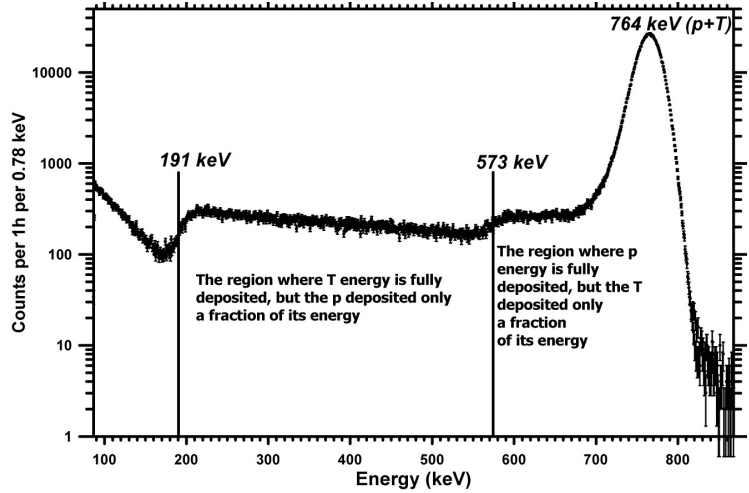}

\noindent\caption{CHM-57 calibration spectrum with exposure to a
strong PuBe source (neutrons from which were moderated by $\sim$2.5~cm of polyethylene).
The vertical lines indicate the kinematic energy
of the T (191\,keV) and p (573\,keV) from the neutron capture on
$^3$He.\label{neutron_spectrum1}}
\end{center}
\end{figure}

The charge from gas-amplified ionization appearing on the signal
wire of the counter is read out by an attached Cremat CR-110 single
channel charge sensitive preamplifier module. It generates a tail
voltage pulse for the shaping amplifier. The spectroscopy amplifier
generates a shaped positive pulse for a 12 bit analog to digital
converter (ADC) with an integrated discriminator. The digitized
pulses are then transferred to a PC by serial connection. An
ORTEC\,448 pulse generator has been used to monitor the electronic
chain. Linearity of the energy scale and the proportionality mode of
the counter has been verified in the JINR laboratory in Dubna using
an intense neutron source. A typical calibration spectrum is
shown in Fig.~\ref{neutron_spectrum1}.
Neutron capture on $^3$He near the counter wall leads to the escape
of one of the two decay products, either triton (T) or proton. This
leads to flanks in the pulse height spectrum as can be seen in
Fig.~\ref{neutron_spectrum1} which can be used to control the
linearity of the energy scale. This was used for estimation of
an uncertainty on the number of counts at energy interval from
660 to 830~keV arising from the energy scale calibration with the single peak.
This uncertainty was found to be below of 0.5\%. The interval from 660 to 830~keV
is the region-of-interest (ROI) used for experimental data analysis
(see section~\ref{section_expt}).

 After installation in LSM, the
position and the stability of the neutron peak in the pulse height
spectrum has been cross-checked with a weak (20~n/sec) AmBe neutron
source.

\section{Detector sensitivity}

The detection sensitivity of the naked $^3$He counter CHM-57 for
thermal neutrons has been determined using the GEANT4
toolkit~\cite{geant4} with the implementation of the GEANT4 Physics
List for low-energy neutron applications QGSP\_BIC\_HP.
For an isotropic ($4\pi$) neutron flux at 1~neutron/cm$^2$/sec the sensitivity was found
to be respectively 243 counts/sec and 227 counts/sec for
2 cases: first one for Maxwell–-Boltzmann distribution $f_{MB}(E)= \frac{E}{(kT_m)^2} \exp{(-\frac{E}{(kT_m)})}$; second one for
the Maxwell–-Boltzmann distribution multiplied by the square root of the energy  $f_{MB}(E) \sqrt{E}$, that could take into account
effect of thermalization, where $T_m=293$~K and $E<0.3$~eV.
Hereinafter for determination of flux of thermal neutrons we will use the
first value as a traditional description of such neutrons. The second
value is needed for the purposes of illustration of systematic
associated with difference of an "ideal" thermal neutron flux and the
"real" one.
The above sensitivities are determined for a pulse height interval corresponding to the 660--830~keV ROI introduced in previous section
(GEANT4 accounts for Proton and Triton energy losses in the counter).
The estimation of detector's sensitivity with GEANT4 has been verified
in calibration measurements with a source of 944.7~g of depleted
$^{238}$U where the 4 CHM-57 counters were placed inside a polyethylene
moderator. The source had been positioned directly on the moderator's surface.
The experimental value of expected neutron flux from 1~kg of natural uranium of 14.9$\pm$0.2~Hz was taken from~\cite{uranium}.
For our source it corresponds to an expected neutron emission rate at
14.2$\pm$0.2~Hz.
The calibration measurements were performed at Dubna laboratory at sea
level, thus we gave special attention to the study of fluctuations of the natural neutron background.
As can be seen in Table~\ref{tabcalib} no significant fluctuations were observed
on the day of measurement.
But since we did observe a fluctuation of measurements performed
on other days, we derived the value for the background as an average between measurements presented
in the table, which is equal to 0.56~Hz.
Two background measurements before and after to one with the
source have respectively $+$0.04 and $-$0.01~Hz differences from the average value.
To take into account possible unstability of ambient neutron background, uncertainty of the average value was
estimated to be 0.04~Hz (as maximal one from two above values).
 The resulting experimental counting rate from the source
is 0.99$\pm$0.06~Hz. GEANT4 predicted count rate of 1.035$\pm$0.014~Hz
(quoted error is from expected source activity only) is in excellent agreement with the measured rate.

\begin{table}[h]
\noindent \caption{Background and calibration runs with $^{238}$U
source performed at Dubna laboratory to check MC prediction of efficiency of CHM-57 counter.
\label{tabcalib}}
\begin{indented}
\item[]\begin{tabular}{@{}lccc}
\br
{\bf Run start time} & {\bf Run time} & {\bf Run type} & {\bf Counting rate at ROI}
            \\
\mr
11h12 & 1200 sec& background & 0.54$\pm$0.02 Hz\\
11h55 & 1000 sec& background & 0.56$\pm$0.02 Hz\\
12h15 & 1000 sec& background & 0.57$\pm$0.02 Hz\\
13h11 & 1000 sec& background & 0.60$\pm$0.02 Hz\\
13h30 & 1000 sec& $^{238}U$+background & 1.55$\pm$0.04 Hz\\
14h22 & 1000 sec& background & 0.55$\pm$0.02 Hz\\
16h12 & 1000 sec& background & 0.58$\pm$0.02 Hz\\
16h32 & 1000 sec& background & 0.54$\pm$0.02 Hz\\
\br
\end{tabular}
\end{indented}
\end{table}

\section{Experimental setup and results\label{section_expt}}

\begin{figure}
\begin{center}
\noindent\epsfxsize=0.29\textwidth \epsfbox{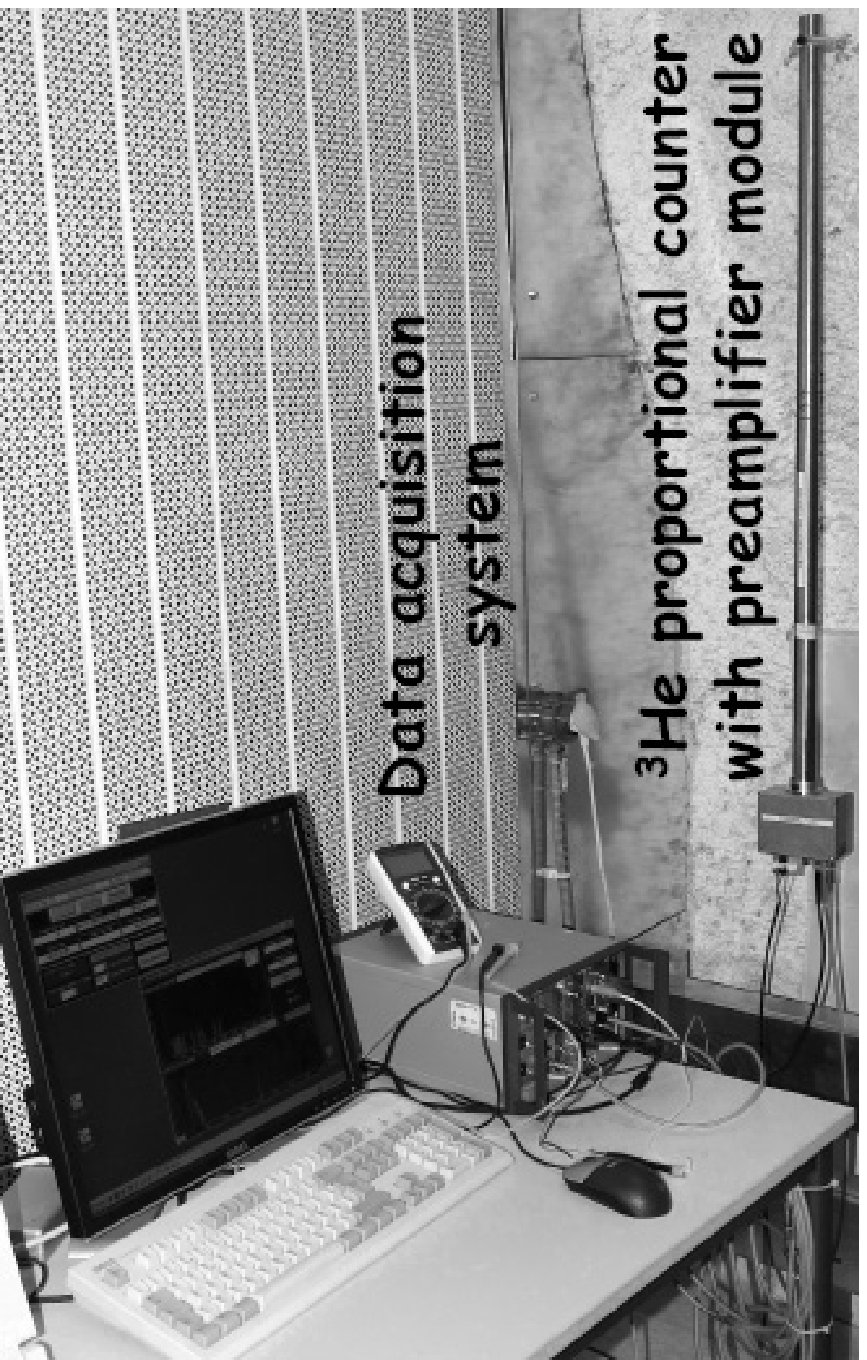}
\epsfxsize=0.70\textwidth \epsfbox{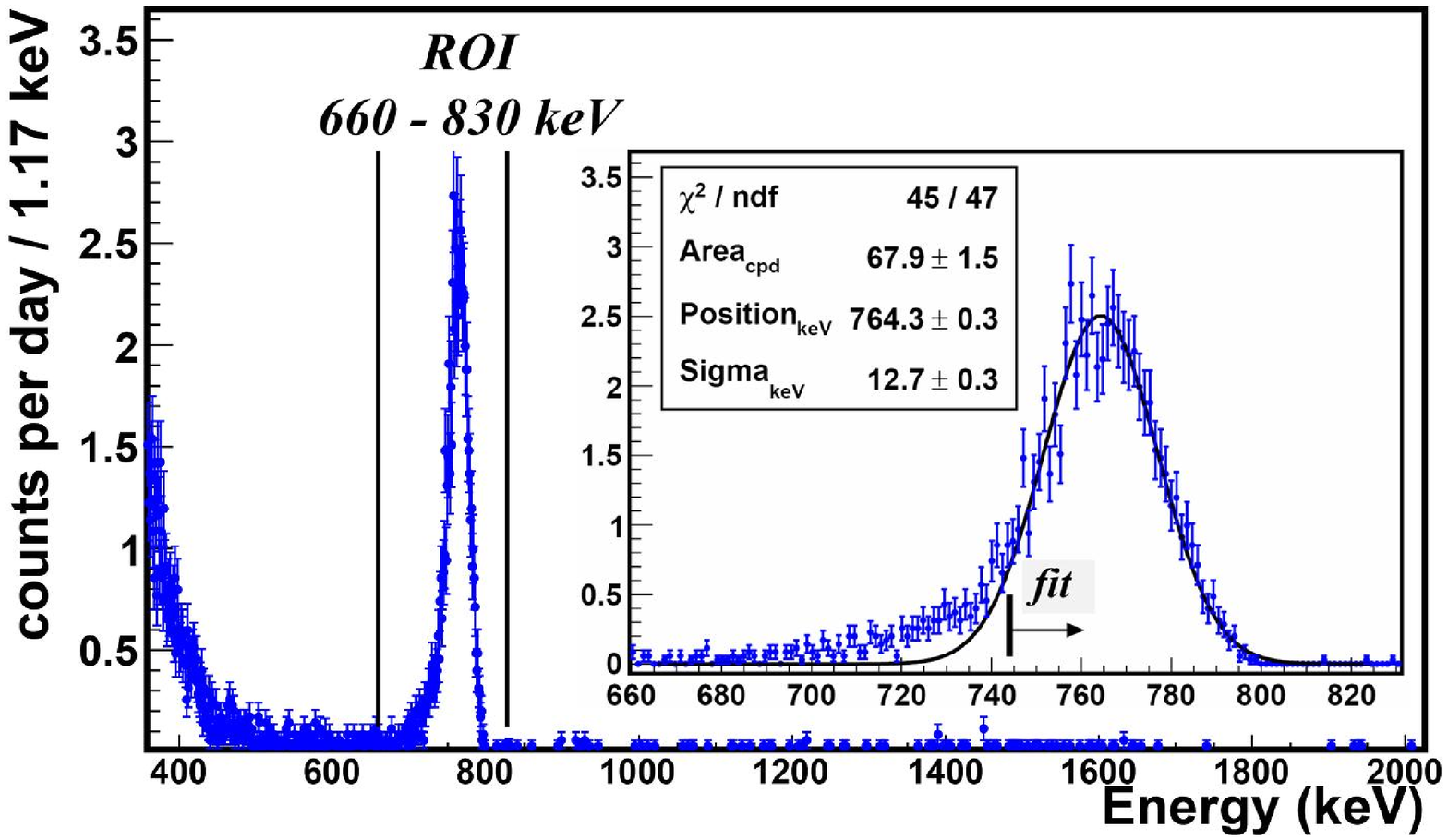}

\noindent\caption{Thermal neutron detection system at
LSM (left) and experimental energy spectrum received for 37 days of measurements (right). Place of
measurement is point 1, see
Fig.~\ref{pic_locations}.\label{neutron_spectrum2}}
\end{center}
\end{figure}

\begin{figure}[h]
\begin{center}
\noindent\epsfxsize=0.7\textwidth \epsfbox{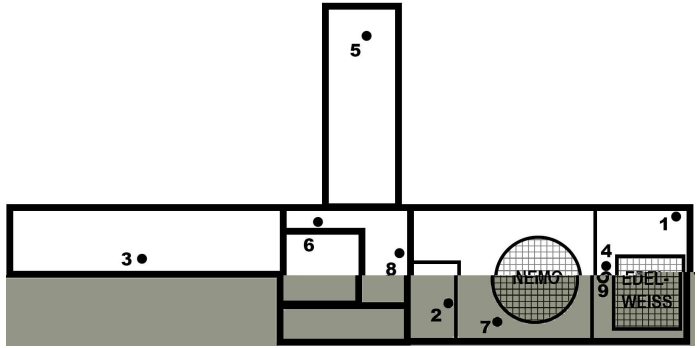}
\noindent\caption{Locations at LSM where measurements of thermal
neutron flux have been performed: 1 -- main place of measurements; 2
-- NEMO preparation clean room; 3~-- LSM tambour; 4 -- between
EDELWEISS and NEMO, approximately same as the point 9 (1992
measurement); 5 -- Ge detectors' hall; 6 -- LSM control room; 7 --
main hall, close to NEMO3 setup; 8 -- platform at the main hall. The
point~9 is the place of the 1992 measurement~\cite{rachid_ref3}.
Places 1, 3, and 8 located on the level 0 of the LSM, all other
points at the level -2.75~m. \label{pic_locations}}
\end{center}
\end{figure}

The thermal neutron monitoring system has been installed
at LSM in November~2008.  The detector was positioned
directly on one of the wall of the laboratory, a few meters
away from the EDELWEISS-II setup. This location is shown on
Figs.~\ref{neutron_spectrum2} and~\ref{pic_locations}. The close
proximity to the wall provides a solid angle of 2$\pi$
for thermal neutrons emerging directly from the wall.
The thermal neutrons coming from the other 2$\pi$ are most likely
affected by materials inside the laboratory, and
especially by the massive polyethylene anti-neutron shield
of the EDELWEISS-II setup.
Such a bias from a pure measurement of the thermal neutron flux
due to the natural radioactivity of the rock
is unavoidable in a fully-operating laboratory like the LSM.
However, the main concern for the EDELWEISS-II experiment
is the actual flux in the laboratory in its present state,
rather than an ideal ``unaffected flux''.

The experimental spectrum recorded from November 4, 2008 to December 10, 2008
(accumulated live time is 35.6 days)
is shown in Fig.~\ref{neutron_spectrum2}. Another run of data taking was performed from
May 27, 2009 to July 27, 2009 (accumulated live time is 52.7
days).
The average rate in
the ROI from 660 to 830 keV for all time of measurement is 75.3$\pm$0.9~cpd. The FWHM resolution of the
detector at 764~keV is 4\%.

The $\alpha$-background in the ROI defined above has been
estimated to be 0.39$\pm$0.04~cpd
from the extrapolation of this background from
the energy region from 0.9 to 1.3~MeV.  This extrapolated value has been
confirmed in direct measurements when the detector has been placed
inside of additional massive polyethylene shields at LSM. In this
study 71 counts were detected in the ROI in 177.8 days of measurements,
with no sign of a neutron peak. This corresponds to an alpha
background rate in the ROI of 0.40$\pm$0.05~cpd, in
agreement with the extrapolated value.
Taking into account the detector sensitivity and the $\alpha$
background, the observed count rate
corresponds to a flux of
$\Phi_{MB}=3.57\pm0.05^{stat}\pm0.27^{syst}\times 10^{-6}$~neutrons/cm$^2$/sec
for this particular location at LSM under the assumption that the neutron spectrum below 0.3~eV is Maxwell-Boltzmann distribution and that
thermal neutron flux is fully isotropic. The systematic error contributions
are listed in Table~\ref{tabsyst}.
Very often it is more useful to define neutron flux
as number of neutrons entering into detector through unit of it
surface area (for example, neutron flux on the surface of EDELWEISS). For this purpose our number for all directional $4\pi$ flux
has to be divided by 2.

We performed additional measurements at various other locations
at LSM, further away from the EDELWEISS-II setup.
The results are
summarized in Table~\ref{tabfluxes}. The point~4
corresponds approximately to the place where a thermal neutron
flux measurement was performed in 1992,
resulting in a flux of
$\Phi=1.6\pm0.1\times 10^{-6}$~neutrons/cm$^2$/sec~\cite{rachid_ref3}.
 This value and the present measurement of
$\Phi_{MB}=2.0\pm 0.2^{stat}\pm 0.15^{syst}\times 10^{-6}$~neutrons/cm$^2$/sec
are in reasonable agreement,
as the former pre-dates the installation of the present
NEMO and EDELWEISS setups.

\begin{table}[h]
\noindent \caption{Thermal neutron fluxes (10$^{-6}$ /cm$^2$/sec) at different locations at
LSM. See Fig.~\ref{pic_locations} for details.\label{tabfluxes}
Errors are statistical only, systematic uncertainties are listed
in Table~\ref{tabsyst}.}
\begin{indented}
\item[]\begin{tabular}{@{}lcccc}
\br
{\bf Point}& {\bf Run time} & \multicolumn{3}{c}{\bf Results} \\\cline{3-5}
           & Days & Counting rate at ROI, cpd& Flux$^1$ & Flux$^2$
            \\
\mr
1$^3$ & 35.6 & 76.8$\pm$1.5 & 3.64$ \pm$0.07  &    3.90$\pm$0.07\\
1$^4$ & 52.7 & 74.3$\pm$1.2 & 3.52$\pm$0.06   & 3.77$\pm$0.06\\
2 & 1.13 & 96.9$\pm$9.3& 4.6$\pm$0.4   & 4.9$\pm$0.5\\
3 & 0.89 & 130.7$\pm$12.1&  6.2$\pm$0.6  &  6.6$\pm$0.6\\
4 & 2.75 & 43.3$\pm$4.0 &   2.0$\pm$0.2 & 2.2$\pm$0.2\\
5 & 1.00 & 94.7$\pm$9.7 &  4.5$\pm$0.5  & 4.8$\pm$0.5\\
6 & 1.17 & 81.8$\pm$8.4&  3.9$\pm$0.4  & 4.1$\pm$0.4\\
7 & 0.91 & 60.4$\pm$8.1&  2.9$\pm$0.4  & 3.1$\pm$0.4\\
8 & 0.83 & 72.2$\pm$9.3 &  3.4$\pm$0.4 &  3.7$\pm$0.5\\
\br
\end{tabular}\\
$^1$ -- Maxwell--Boltzmann distribution\\
$^2$ -- Maxwell--Boltzmann multiplied by $\sqrt{E}$\\
$^3$ -- measurements performed from November 4, 2008 to December 10,
2008\\
$^4$ -- measurements performed from June 1, 2009 to July 28, 2009
\end{indented}
\end{table}

\begin{table}[h]
\noindent \caption{Estimated systematic uncertainties (\%).\label{tabsyst}}
\begin{indented}
\item[]\begin{tabular}{@{}llll}
\mr
Active volume     & 4.6 & Detector's sensitivity (MC) & 6.2\\
Background        & $<$0.5 & ROI determination & $<$0.5\\
Live time         & $<$0.5 &  \\
\multicolumn{3}{@{}l}{Total systematic error} & 7.7\%\\
\br
\end{tabular}\\

\end{indented}
\end{table}

\begin{figure}
\begin{center}
\noindent\epsfxsize=0.475\textwidth \epsfbox{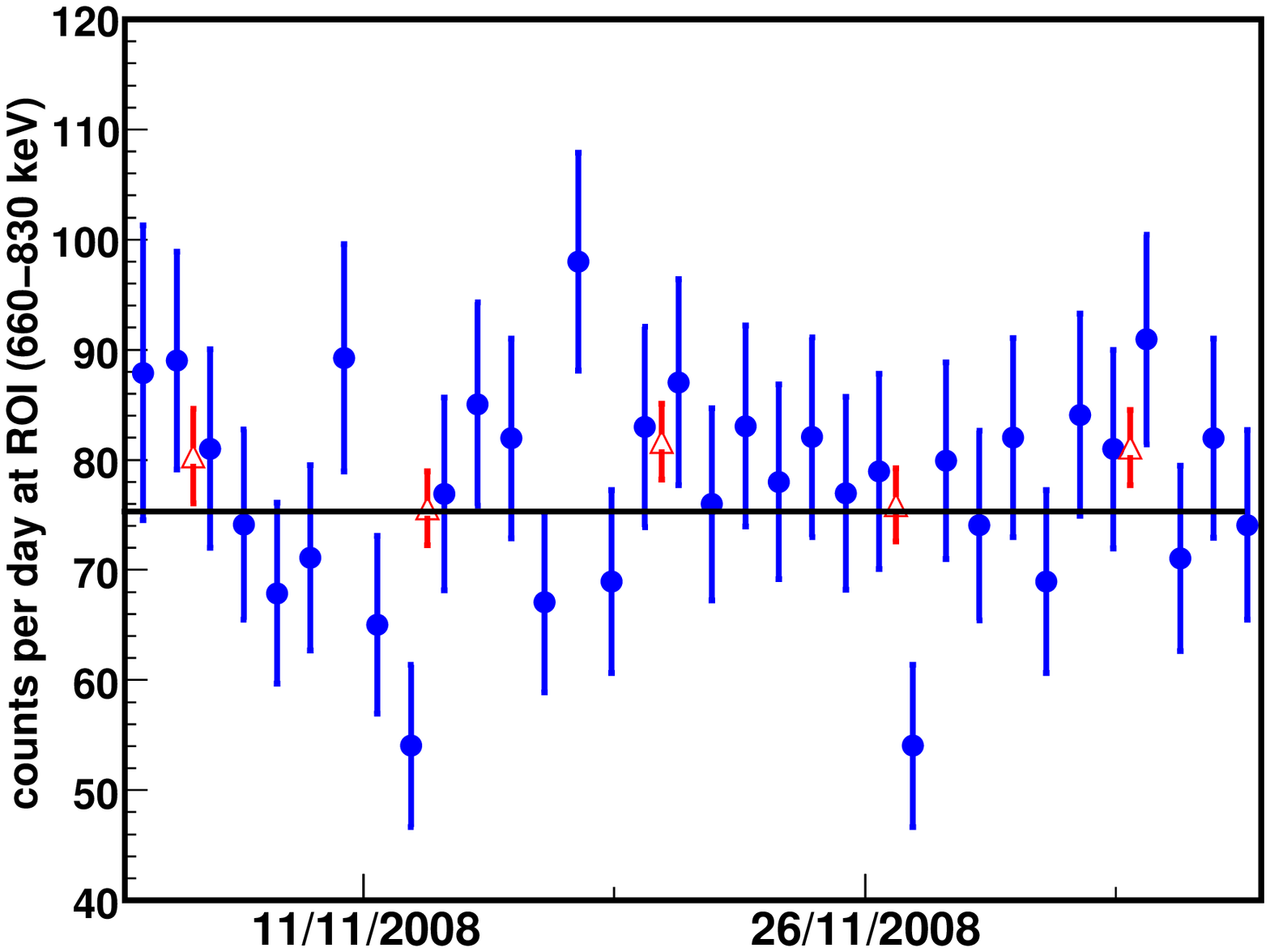}
\epsfxsize=0.475\textwidth \epsfbox{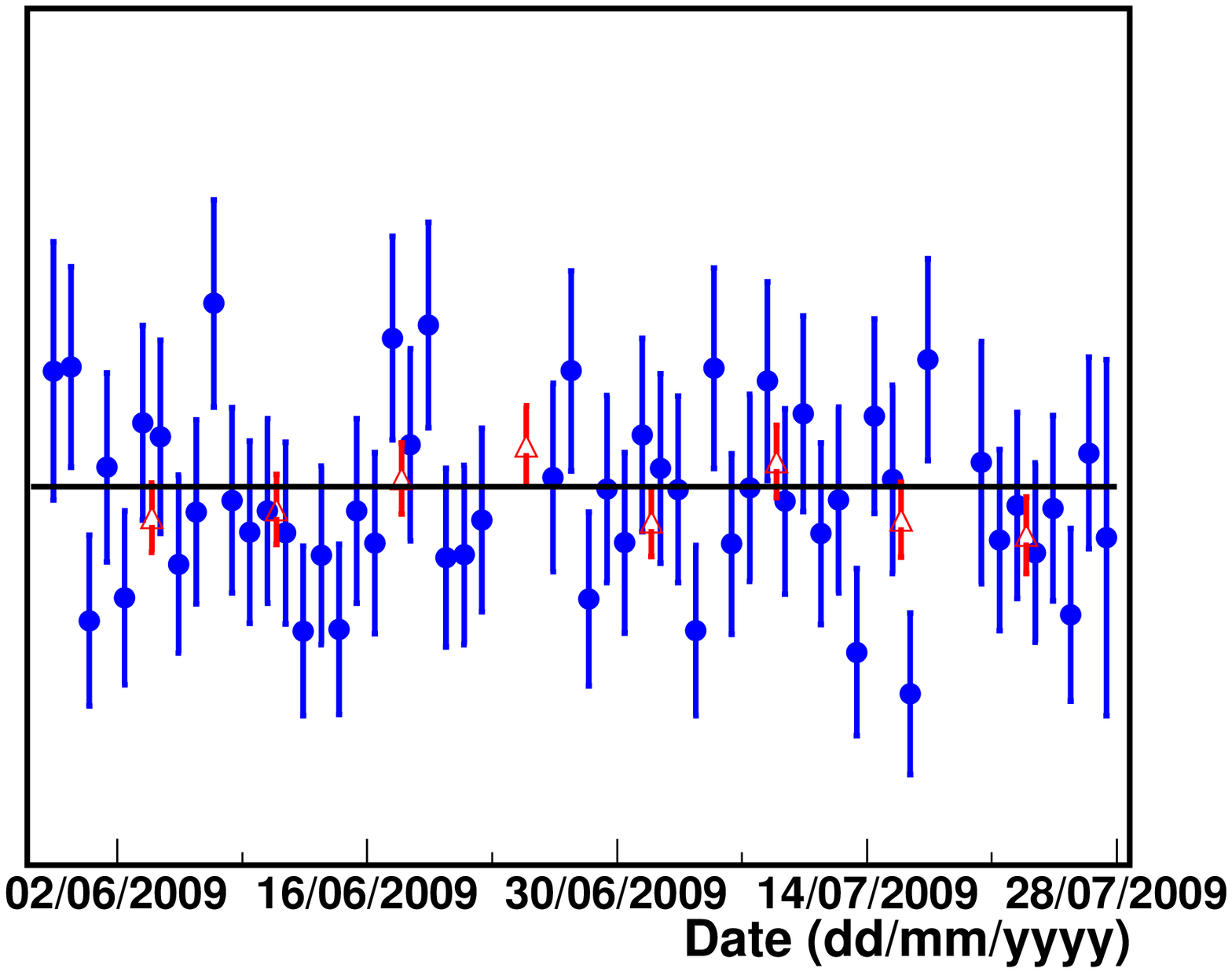}

\noindent\epsfxsize=0.95\textwidth \epsfbox{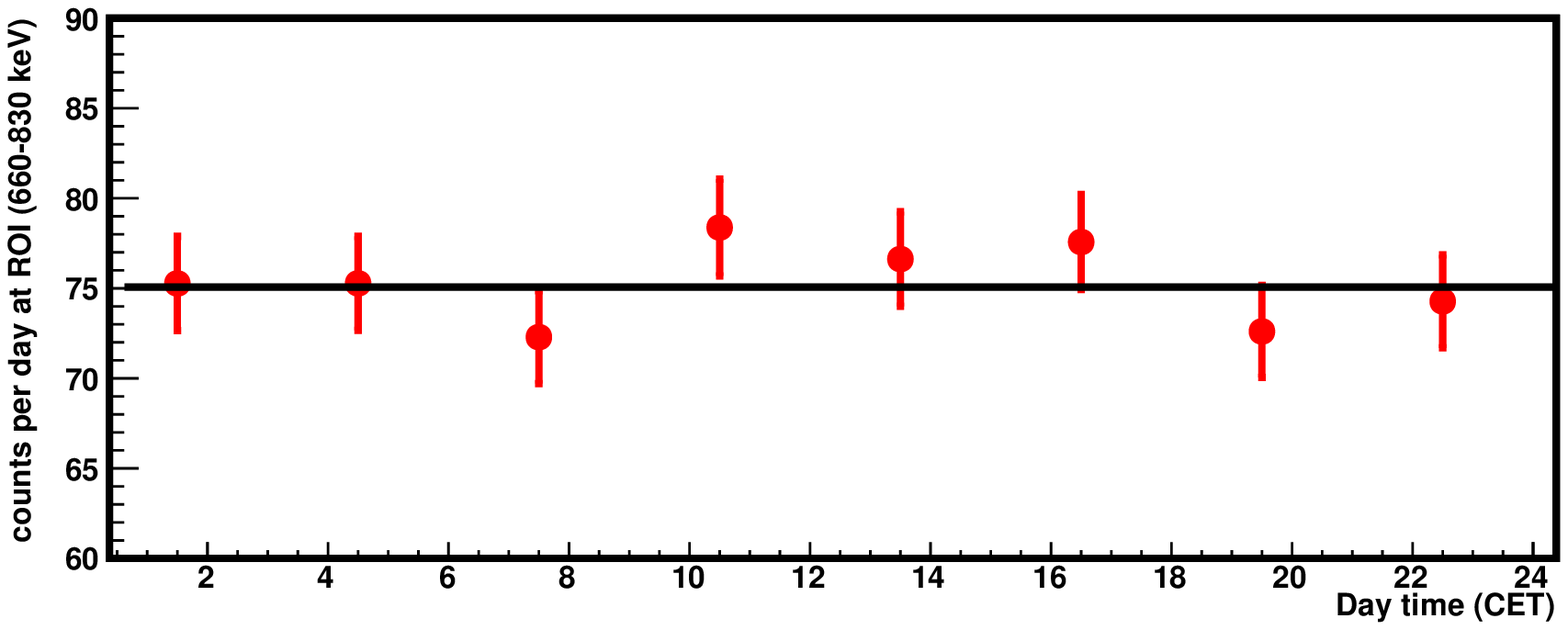}

\noindent\caption{Changes of neutron flux at LSM with time. On the upper plot, points correspond to day by day sampling intervals,
triangles correspond to week by week sampling (with weeks starting on Sunday).
On the bottom plot, points correspond to particular day time with 3 h step.
Lines on all plots are average counts' rate for all time of
measurement.
\label{fig_time_change_lsm}}
\end{center}
\end{figure}

Table~\ref{tabfluxes} clearly shows that the thermal neutron flux
at LSM may vary by up to a factor three from one location to
an other.
In addition to the large neutron shields, additional
point-to-point flux variations may be due to other materials present
at LSM, and in particular their water content,
as well as to inhomogeneities in material and their U/Th
contamination.
The high efficiency of the neutron detector and the resulting
large count rate, as well as the low background rate,
make a detailed study of the
variation in time of the thermal neutron flux possible, as shown
in Fig.~\ref{fig_time_change_lsm} and Table~\ref{tabfluxes1}. All observed fluctuations are in
agreement with statistical expectation.

\begin{table}[h]
\noindent \caption{Statistical analysis of time variations of measured neutron flux.\label{tabfluxes1}}
\begin{indented}
\item[]\begin{tabular}{@{}lccc}
\br
{\bf Sampling method }& {\bf $\chi^2$/ndf}$^*$ & {\bf $\chi^2$/ndf$^+$  90\% CL} & {\bf $\sigma$ at single sample} \\
\mr
day / 3h intervals & 4.9/7$=$0.7 & 0.5$<\chi^2<$1.7 & $\sim$3.5\%
\\
day by day & 96/91$=$1.05 & 0.8$<\chi^2<$1.2 & $\sim$11\%\\
week by week & 15.6/15$=$1.04 &  $0.6<\chi^2<1.5$ & $\sim$4\%\\
\br
\end{tabular}\\
$^*$ - fit of experimental data with the assumption that neutron flux is constant
with time\\
$^+$ -    both tail's areas of $\chi^2$ distribution (below and above of the interval) are 10\% \\
\end{indented}

\end{table}

\section{Conclusion}

As part of the program of background controls for the EDELWEISS-II
experiment environment, the thermal neutron
flux is monitored using a $^{3}$He filled bare proportional counter.
With a daily count rate of 75 and a background of a fraction of
count, this detector makes possible a
continuous day by day monitoring of the thermal
neutron flux. No departure from a constant flux with time was measured on scales of
3h/day/week, at a given location. However, variations of the
thermal flux as large a factor three were measured between different
locations in the laboratory.
The flux close to the laboratory wall at the proximity of the
EDELWEISS-II setup is
$\Phi_{MB}=3.57\pm0.05^{stat}\pm0.27^{syst}\times 10^{-6}$~neutrons/cm$^2$/sec
under the assumption that the neutron spectrum follows a Maxwell-Boltzmann distribution and that
thermal neutron flux is isotropic.

\section{Acknowledgements}

The EDELWEISS collaboration is grateful to the LSM laboratory  that
provided service for work underground and highly valuable technical
assistance. We would like to give special recognition to
B.~Branlard for his valuable contribution to works performed at LSM.
We are obliged to Dr.~Yu.~Shitov for his experienced
help with Geant4 simulations. This work has been partly supported by
Russian Foundation for Basic Research (grant~No.~07-02-00355-a) and
ILIAS (TARI project P2008-01-LSM).

\end{document}